\documentclass[english,aip]{revtex4-1}
\usepackage[T1]{fontenc}
\usepackage[utf8]{inputenc}
\usepackage{array}
\usepackage{multirow}
\usepackage{amsmath}
\usepackage{graphicx}

\makeatletter

\newcommand{\noun}[1]{\textsc{#1}}
\providecommand{\tabularnewline}{\\}

\usepackage{mhchem}
\usepackage{hyperref}

\makeatother

\usepackage{babel}
\begin{document}
\title{Molecular vibrational frequencies from analytic Hessian of constrained
nuclear-electronic orbital density functional theory}
\author{Xi Xu}
\author{Yang Yang}
\email{yyang222@wisc.edu}

\affiliation{Theoretical Chemistry Institute and Department of Chemistry, University
of Wisconsin-Madison, 1101 University Avenue, Madison, WI 53706, USA}
\date{\today}
\begin{abstract}
Nuclear quantum effects are important in a variety of chemical and
biological processes. The constrained nuclear-electronic orbital density
functional theory (cNEO-DFT) has been developed to include nuclear
quantum effects in energy surfaces. Herein we develop the analytic
Hessian for cNEO-DFT energy with respect to the change of nuclear
(expectation) positions, which can be used to characterize stationary
points on energy surfaces and compute molecular vibrational frequencies.
This is achieved by constructing and solving the multicomponent cNEO
coupled-perturbed Kohn-Sham (cNEO-CPKS) equations, which describe
the response of electronic and nuclear orbitals to the displacement
of nuclear (expectation) positions. With the analytic Hessian, the
vibrational frequencies of a series of small molecules are calculated
and compared to those from conventional DFT Hessian calculations as
well as those from the vibrational second-order perturbation theory
(VPT2). It is found that even with a harmonic treatment, cNEO-DFT
significantly outperforms DFT and is comparable to DFT-VPT2 in the
description of vibrational frequencies in regular polyatomic molecules.
Furthermore, cNEO-DFT can reasonably describe the proton transfer
modes in systems with a shared proton, whereas DFT-VPT2 often faces
great challenges. Our results suggest the importance of nuclear quantum
effects in molecular vibrations, and cNEO-DFT is an accurate and inexpensive
method to describe molecular vibrations.
\end{abstract}
\maketitle

\section{Introduction}

Multicomponent quantum theory has been an emerging research field
in quantum chemistry.\citep{Kreibich2001,Bochevarov2004,Nakai2007a,Ishimoto2009,Abedi2010,Pavosevic2020}
It simultaneously treats at least two types of particles, such as
electrons and nuclei, quantum-mechanically and avoids the conventional
Born-Oppenheimer (BO) approximation between electrons and nuclei.
Thereby, the nuclear quantum effects, which are crucial in many hydrogen-bonded
systems such as water,\citep{Ceriotti2016,Guo2017} can be described
together with electronic quantum effects. The nuclear-electronic orbital
(NEO) framework\citep{Webb2002,Pavosevic2020} is a simple and popular
type of multicomponent quantum theory. It simultaneously treats electrons
and key nuclei, typically protons, within the orbital picture. In
the NEO framework, both wave-function-based methods and density functional
theory have been developed, and they have been successful in describing
the ground and excited state properties of many small molecules.\citep{Pavosevic2020}
However, in conventional NEO calculations, at least two nuclei have
to be treated classically to fix the molecular frame and avoid the
problems related to translations and rotations.\citep{Iordanov2003}
This practice introduces a new BO separation between the classical
and quantum nuclei, which assumes the quantum nuclei respond instantaneously
to the motion of classical nuclei. It makes neither the vibrational
excitations from NEO time-dependent density functional theory (NEO-TDDFT)
nor the normal modes from NEO Hessian directly correspond to the vibrations
observed in experiments.\citep{Iordanov2003,Yang2018a,Yang2019,Schneider2021}

Recently, we developed constrained NEO-DFT (cNEO-DFT) to overcome
the challenge. In cNEO-DFT, we introduce constraints on the expectation
values of the quantum nuclear positions and minimize the total energy
under the constraints. These constrained expectation positions naturally
fix the molecular frame and therefore enable a full-quantum treatment
of molecules. The full-quantum treatment avoids any BO approximation.\citep{Xu2020a}
In cNEO theory, the energy surface is a function of the quantum nuclear
expectation positions as well as the classical nuclear positions.\citep{Xu2020}
This is similar to BO potential energy surface, but because quantum
nuclei are described by orbitals rather than fixed point charges,
the cNEO energy surfaces incorporate nuclear quantum effects, in particular
the zero-point effects. Previously, we have successfully performed
geometry optimizations and transition states search on the cNEO energy
surfaces for several simple molecular systems and chemical reactions.\citep{Xu2020a}
The zero-point effects and geometric isotope effects have been observed
on the cNEO surface, suggesting the promising future of cNEO-DFT in
describing systems and reactions with significant nuclear quantum
effects. However, the characterization of reactants, products, and
transition states on the cNEO surface is needed when applying cNEO
theory to practical quantum chemistry calculations, which requires
the computation of the Hessian of cNEO energies. In this work, we
will derive and implement the analytic Hessian for cNEO energies and
investigate its performance through molecular vibrational frequency
calculations.

The Hessian can be obtained either numerically or analytically. However,
the numerical Hessian is often more expensive and can suffer from
the amplification of numerical error or the contamination of higher-order
terms.\citep{Pulay2014} In contrast, analytic Hessian has higher
computational efficiency and numerical accuracy.\citep{Pulay2014}
In conventional electronic structure theory, the calculation of analytic
Hessian requires the response of the electronic orbitals to the perturbative
nuclear displacement, which can be obtained by solving coupled perturbed
Hartree-Fock/Kohn-Sham equations.\citep{Pople1979,Pulay1983,Komornicki1993,Yamaguchi2011}
In cNEO-DFT, the analytic Hessian requires the response of both electronic
and nuclear orbitals to the perturbative change in quantum nuclear
expectation positions as well as classical nuclear positions. It requires
solving a multicomponent cNEO coupled perturbed Kohn-Sham (cNEO-CPKS)
equation, which couples the response of electrons and quantum nuclei.
In addition, the constraints in nuclear expectation positions lead
to additional terms incorporated in the cNEO-CPKS equation.

The rest of the paper is organized as follows. In Sec. II, we present
the expression of analytic Hessian for cNEO-DFT as well as the formulation
of cNEO-CPKS equations. Then we investigate the vibrational frequencies
of a series of molecules by building and diagonalizing the cNEO Hessian
matrices, with the computational details presented in Sec III and
results presented in Sec IV. The performance of cNEO-DFT is further
compared with other commonly used methods, including harmonic DFT,
DFT-VPT2, and NEO-DFT(V). We give our concluding remarks in Sec V.

\section{Theory}

\subsection{Energy and derivatives}

In cNEO-DFT, the electrons can be treated in the same way as in conventional
DFT, whereas quantum nuclei are treated as distinguishable particles
because they are relatively localized in space.\citep{Xu2020,Xu2020a}
The total energy can be written as

\begin{align}
E & =\sum_{i}h_{ii}+\frac{1}{2}\sum_{ij}(ii|jj)+\sum_{n}h_{I_{n}I_{n}}+\frac{1}{2}\sum_{n,n'}Z_{n}Z_{n'}(I_{n}I_{n}|I_{n'}I_{n'})-\sum_{n}\sum_{i}Z_{n}(ii|I_{n}I_{n})\nonumber \\
 & +E^{\text{xc}}[\rho^{e},\{\rho^{n}\}]+\frac{1}{2}\sum_{A}\sum_{B\neq A}\frac{Z_{A}Z_{B}}{|\mathbf{R}_{A}-\mathbf{R}_{B}|},
\end{align}
where each term represents the core Hamiltonian for electrons, the
electronic Coulomb interaction, the core Hamiltonian for nuclei, the
nuclear Coulomb interaction, the electron-nuclear Coulomb interaction,
the exchange-correlation energy as a functional of both electronic
density $\rho^{e}$ and nuclear densities $\{\rho^{n}\}$, and the
Coulomb repulsion between classical nuclei. In this paper, we will
use $i,j,k,l$ to denote occupied molecular orbitals, $a,b,c,d$ to
denote unoccupied molecular orbitals, and $p,q,r,s,m$ to denote general
molecular orbitals. We will use the corresponding upper case letters
to denote nuclear orbitals. Because quantum nuclei are treated as
distinguishable particles, each nucleus occupies one orbital and we
use $I_{n}$ ( or $J_{n}$, $K_{n}$) to denote the only occupied
orbital for the $n$th quantum nucleus. We will use $\mu,\nu$ to
denote electronic atomic orbitals and $\mu_{n},\nu_{n}$ to denote
the atomic orbitals for the $n$th quantum nucleus. The exchange-correlation
energy $E^{\text{xc}}$ includes electronic exchange-correlation,
nuclear exchange-correlation, and electronic-nuclear correlation.
The definition of the two-particle integral is

\begin{equation}
(ij|kl)=\int\phi_{i}(\mathbf{r}')\phi_{j}(\mathbf{r}')\frac{1}{|\mathbf{r}-\mathbf{r}'|}\phi_{k}(\mathbf{r})\phi_{l}(\mathbf{\mathbf{r}})\mathrm{d}\mathbf{r}'\mathrm{d}\mathbf{r}.
\end{equation}

The constraint on the expectation position of the $n$th quantum nucleus
is
\begin{equation}
\mathbf{r}_{I_{n}I_{n}}\equiv(I_{n}|\mathbf{r}|I_{n})\equiv\int\phi_{I_{n}}(\mathbf{r})\mathbf{r}\phi_{I_{n}}(\mathbf{\mathbf{r}})\mathrm{d}\mathbf{r}=\mathbf{R}_{n},\label{eq:constraint}
\end{equation}
and the additional term in the Lagrangian for the constraint is
\begin{equation}
\sum_{n}\mathbf{f}_{n}\cdot(\mathbf{r}_{I_{n}I_{n}}-\mathbf{R}_{n}),
\end{equation}
where $\mathbf{f}_{n}$ is the Lagrange multiplier for the $n$th
quantum nucleus, and it has been proven to be the classical force
acting on the nucleus in the complete basis set limit.\citep{Xu2020}

Minimizing the Lagrangian with respect to electronic and nuclear densities
leads to the Fock equations, and the electronic and nuclear Fock matrix
elements are defined as
\begin{equation}
F_{pq}=h_{pq}+\sum_{k}(pq|kk)-\sum_{n}Z_{n}(pq|I_{n}I_{n})+V_{pq}^{\text{xc}},\label{eq:fock_elec}
\end{equation}
and

\begin{equation}
F_{P_{n}Q_{n}}=h_{P_{n}Q_{n}}-\sum_{k}Z_{n}(P_{n}Q_{n}|kk)+\sum_{n'}Z_{n}Z_{n'}(P_{n}Q_{n}|I_{n'}I_{n'})+V_{P_{n}Q_{n}}^{\text{xc}}+\mathbf{f}_{n}\cdot(P_{n}|\mathbf{r}|Q_{n}),\label{eq:fock_nuc}
\end{equation}
respectively, where $V^{\text{xc}}$ is the exchange-correlation potential
for either electrons or nuclei depending on the subscripts.

The gradient and Hessian of the energy requires the computation of
the derivatives for some key components, including $h_{ii}$, $(ii|jj)$,
$h_{I_{n}I_{n}}$, $(I_{n}I_{n}|I_{n'}I_{n'})$, $(ii|I_{n}I_{n})$,
$E^{\text{xc}}[\rho^{e},\{\rho^{n}\}]$, which can be obtained analogously
to conventional electronic Hartree-Fock and Kohn-Sham DFT. The detailed
derivation for the required terms are presented in the Supplementary
Materials. However, unlike conventional theories with only constraints
on orbital normalization, cNEO-DFT additionally imposes the constraints
on nuclear expectation positions and therefore requires the derivative
of the constraints in the derivation of gradient and Hessian. The
derivative of the constraint (Eq. \ref{eq:constraint}) with respect
to a perturbation $\xi$ is
\begin{equation}
\frac{\partial\mathbf{R}_{n}}{\partial\xi}=\mathbf{r}_{I_{n}I_{n}}^{\xi}+\sum_{M_{n}}(\mathbf{r}_{I_{n}M_{n}}U_{M_{n}I_{n}}^{\xi}+U_{M_{n}I_{n}}^{\xi}\mathbf{r}_{M_{n}I_{n}}),\label{eq:constraint_de}
\end{equation}
where the matrix $\mathbf{U}^{\xi}$ follows the same definition as
Ref. \onlinecite{Yamaguchi2011} and describes the response of the
molecular orbital coefficient matrix $\mathbf{C}$ to the perturbation
$\xi$: ,
\begin{align}
\frac{\partial C_{\mu_{n}P_{n}}}{\partial\xi} & =\sum_{M_{n}}C_{\mu_{n}M_{n}}U_{M_{n}P_{n}}^{\xi},
\end{align}
where $\mathbf{r}_{I_{n}I_{n}}^{\xi}$ is defined as
\begin{equation}
\mathbf{r}_{I_{n}I_{n}}^{\xi}=\sum_{\mu_{n}\nu_{n}}C_{\mu_{n}I_{n}}C_{\nu_{n}I_{n}}\frac{\partial\mathbf{r}_{\mu_{n}\nu_{n}}}{\partial\xi}.\label{eq:supersciptdef}
\end{equation}

The first-order derivative for the total energy is
\begin{align}
\frac{\partial E}{\partial\xi} & =\sum_{i}h_{ii}^{\xi}+\frac{1}{2}\sum_{ij}(ii|jj)^{\xi}+\sum_{n}h_{I_{n}I_{n}}^{\xi}-\sum_{n}\sum_{i}Z_{n}(ii|I_{n}I_{n})^{\xi}+\frac{1}{2}\sum_{n,n'}Z_{n}Z_{n'}(I_{n}I_{n}|I_{n'}I_{n'})^{\xi}\nonumber \\
 & +\sum_{i}V_{ii}^{\text{\text{xc}},\xi}+\sum_{n}V_{I_{n}I_{n}}^{\text{xc},\xi}-\sum_{i}\epsilon_{i}S_{ii}^{\xi}-\sum_{n}\epsilon_{I_{n}}S_{I_{n}I_{n}}^{\xi}+\sum_{n}\mathbf{f}_{n}\cdot\mathbf{r}_{I_{n}I_{n}}^{\xi}-\sum_{n}\mathbf{f}_{n}\cdot\frac{\partial\mathbf{R}_{n}}{\partial\xi},\label{eq:gradient}
\end{align}
where $\epsilon$ is the orbital energy, $S$ is the overlap matrix,
and quantities with the superscript $\xi$ are partial derivatives
defined analogously to Eq. \ref{eq:supersciptdef} in which the molecular
orbital coefficients $\mathbf{C}$ is not differentiated. When $\xi$
is chosen to be the expectation position of a particular quantum nucleus
$\xi=\mathbf{R}_{n_{0}}$, this expression is the same as the gradient
result in our previous work,\citep{Xu2020a} with the last term being
\begin{equation}
-\sum_{n}\mathbf{f}_{n}\cdot\frac{\partial\mathbf{R}_{n}}{\partial\mathbf{R}_{n_{0}}}=-\mathbf{f}_{n_{0}}.
\end{equation}

The second-order derivative for the total energy can be obtained by
taking the derivative of Eq. \ref{eq:gradient} with respect to another
perturbation $\chi$. The final expression is 
\begin{align}
\frac{\partial^{2}E}{\partial\xi\partial\chi} & =\sum_{i}h_{ii}^{\xi\chi}+\frac{1}{2}\sum_{ij}(ii|jj)^{\xi\chi}+\sum_{n}h_{I_{n}I_{n}}^{\xi\chi}-\sum_{i,n}Z_{n}(ii|I_{n}I_{n})^{\xi\chi}+\frac{1}{2}\sum_{n,n'}Z_{n}Z_{n'}(I_{n}I_{n}|I_{n'}I_{n'})^{\xi\chi}\nonumber \\
 & +\sum_{i}V_{ii}^{\text{xc},\xi\chi}+\sum_{n}V_{I_{n}I_{n}}^{\text{xc},\xi\chi}\nonumber \\
 & +\sum_{ik}K_{(ii)^{\xi},(kk)^{\chi}}^{\text{xc}}+\sum_{in}K_{(ii)^{\xi},(K_{n}K_{n})^{\chi}}^{\text{xc}}+\sum_{nn'}K_{(I_{n}I_{n})^{\xi},(K_{n'}K_{n'})^{\chi}}^{\text{xc}}+\sum_{nk}K_{(I_{n}I_{n})^{\xi},(kk)^{\chi}}^{\text{xc}}\nonumber \\
 & +2\sum_{i}\sum_{p}(F_{ip}^{\xi}U_{pi}^{\chi}+F_{ip}^{\chi}U_{pi}^{\xi})+2\sum_{n}\sum_{P_{n}}(F_{I_{n}P_{n}}^{\xi}U_{P_{n}I_{n}}^{\chi}+F_{I_{n}P_{n}}^{\chi}U_{P_{n}I_{n}}^{\xi})-\sum_{i}\epsilon_{i}S_{ii}^{\xi\chi}-\sum_{n}\epsilon_{I_{n}}S_{I_{n}I_{n}}^{\xi\chi}\nonumber \\
 & +2\sum_{i}\sum_{p}U_{pi}^{\chi}U_{pi}^{\xi}\epsilon_{p}+2\sum_{n}\sum_{P_{n}}U_{P_{n}I_{n}}^{\chi}U_{P_{n}I_{n}}^{\xi}\epsilon_{P_{n}}\nonumber \\
 & -2\sum_{i}\sum_{p}\epsilon_{i}(U_{ip}^{\chi}U_{ip}^{\xi}-S_{ip}^{\xi}S_{pi}^{\chi})-2\sum_{n}\sum_{P_{n}}\epsilon_{I_{n}}(U_{I_{n}P_{n}}^{\chi}U_{I_{n}P_{n}}^{\xi}-S_{I_{n}P_{n}}^{\xi}S_{P_{n}I_{n}}^{\chi})\nonumber \\
 & +2\sum_{i}\sum_{p}[\sum_{k}\sum_{q}(2(ip|kq)+2K_{ip,kq}^{\text{xc}})U_{qk}^{\chi}U_{pi}^{\xi}-\sum_{n}\sum_{P_{n}}2(Z_{n}(ip|K_{n}P_{n})-K_{ip,K_{n}P_{n}}^{\text{xc}})U_{P_{n}K_{n}}^{\chi}U_{pi}^{\xi}]\nonumber \\
 & +2\sum_{n}\sum_{P_{n}}[\sum_{n'}\sum_{Q_{n'}}(2Z_{n}Z_{n'}(I_{n}P_{n}|K_{n'}Q_{n'})+2K_{I_{n}P_{n},K_{n'}Q_{n'}}^{\text{xc}})U_{Q_{n'}K_{n'}}^{\chi}U_{P_{n}I_{n}}^{\xi}\nonumber \\
 & -\sum_{k}\sum_{p}2(Z_{n}(I_{n}P_{n}|kp)-K_{I_{n}P_{n},kp}^{\text{xc}})U_{pk}^{\chi}U_{P_{n}I_{n}}^{\xi}]\nonumber \\
 & +\sum_{n}\mathbf{f}_{n}\cdot\mathbf{r}_{I_{n}I_{n}}^{\xi\chi}-\mathbf{f}_{n}\cdot\frac{\partial^{2}\mathbf{R}_{n}}{\partial\xi\partial\chi},\label{eq:energy_2de}
\end{align}
where $K^{\text{xc}}$ is the exchange-correlation kernel and the
detailed derivation for the equation is provided in Supporting Information.
The key components in the expression can be calculated analogously
to conventional electronic CPKS. The knowledge of $\mathbf{U}$ can
be obtained by solving the cNEO-CPKS equation, which will be presented
in next section. Note that when $\xi$ and $\chi$ are both chosen
to be the nuclear expectation positions, the last term $-\mathbf{f}_{n}\cdot\frac{\partial^{2}\mathbf{R}_{n}}{\partial\xi\partial\chi}$
vanishes.

\subsection{cNEO-CPKS}

As with conventional electronic DFT, a converged SCF solution for
cNEO-DFT always satisfies $F_{ia}=0$, which leads to 
\begin{equation}
\frac{\partial F_{ai}}{\partial\xi}=0.
\end{equation}
We can calculate the derivative of electronic Fock matrix elements
in a similar way to conventional electronic DFT, but with additional
Coulomb and correlation terms from nuclei.

\begin{align}
0= & \frac{\partial F_{ai}}{\partial\xi}\label{eq:fock_elec_de}\\
= & F_{ai}^{\xi}+(\epsilon_{a}-\epsilon_{i})U_{ai}^{\xi}-S_{ai}^{\xi}\epsilon_{i}+2\sum_{jb}((ai|bj)+K_{ai,bj}^{\text{xc}})U_{bj}^{\xi}-2\sum_{n}\sum_{B_{n}}(Z_{n}(ai|B_{n}J_{n})-K_{ai,B_{n}J_{n}}^{\text{xc}})U_{B_{n}J_{n}}^{\xi}\nonumber \\
 & -\sum_{jk}((ai|kj)+K_{ai,kj}^{\text{xc}})S_{kj}^{\xi}+\sum_{n}(Z_{n}(ai|J_{n}J_{n})-K_{ai,J_{n}J_{n}}^{\text{xc}})S_{J_{n}J_{n}}^{\xi}.\nonumber 
\end{align}
Analogously, the derivative of the nuclear Fock matrix can be derived.
The difference is that the extra Lagrange multiplier term in the nuclear
Fock matrix (Eq. \ref{eq:fock_nuc}) gives an additional term in its
derivative,
\begin{align}
0= & \frac{\partial F_{A_{n}I_{n}}}{\partial\xi}\label{eq:fock_nuc_de}\\
= & F_{A_{n}I_{n}}^{\xi}+(\epsilon_{A_{n}}-\epsilon_{I_{n}})U_{A_{n}I_{n}}^{\xi}-S_{A_{n}I_{n}}^{\xi}\epsilon_{I_{n}}\nonumber \\
 & +2\sum_{n'}\sum_{B_{n'}}(Z_{n}Z_{n'}(A_{n}I_{n}|B_{n'}J_{n'})+K_{A_{n}I_{n},B_{n'}J_{n'}}^{\text{xc}})U_{B_{n'}J_{n'}}^{\xi}-2\sum_{j}\sum_{b}(Z_{n}(A_{n}I_{n}|bj)-K_{A_{n}I_{n},bj}^{\text{xc}})U_{bj}^{\xi}\nonumber \\
 & -\sum_{n'}(Z_{n}Z_{n'}(A_{n}I_{n}|J_{n'}J_{n'})+K_{A_{n}I_{n},J_{n'}J_{n'}}^{\text{xc}})S_{J_{n'}J_{n'}}^{\xi}+\sum_{j}\sum_{k}(Z_{n}(A_{n}I_{n}|kj)-K_{A_{n}I_{n},kj}^{\text{xc}})S_{jk}^{\xi}+\frac{\partial\mathbf{f}_{n}}{\partial\xi}\cdot\mathbf{r}_{A_{n}I_{n}}.\nonumber 
\end{align}
The unknown variables in Eq. \ref{eq:fock_elec_de} and \ref{eq:fock_nuc_de}
are the electronic and nuclear $\mathbf{U}^{\xi}$ matrices as well
as the derivative of $\mathbf{f}$ with respect to $\xi$. These two
sets of equations are not sufficient to solve all unknown variables,
and the derivative of the constraint also needs to be included. Rearranging
Eq. \ref{eq:constraint_de} leads to 
\begin{align}
\sum_{B_{n}}\mathbf{r}_{I_{n}B_{n}}U_{B_{n}I_{n}}^{\xi} & =\frac{1}{2}(\frac{\partial\mathbf{R}_{n}}{\partial\xi}-\mathbf{r}_{I_{n}I_{n}}^{\xi}+\mathbf{R}_{n}S_{I_{n}I_{n}}^{\xi}).\label{eq:constraint_de_rearange}
\end{align}
Note that here we have used the relationship $2U_{I_{n}I_{n}}^{\xi}=-S_{I_{n}I_{n}}^{\xi}$,
which is derived from the derivative of the normalization constraint.\citep{Pople1979}
The three sets of equations Eq. \ref{eq:fock_elec_de}, \ref{eq:fock_nuc_de}
and \ref{eq:constraint_de_rearange} can be cast into a coupled form,
which is the cNEO-CPKS equation,
\begin{equation}
\left[\begin{array}{ccc}
\mathbf{A}^{ee} & \mathbf{A}^{en} & \mathbf{0}\\
\mathbf{A}^{ne} & \mathbf{A}^{nn} & \mathbf{r}\\
\mathbf{0} & \mathbf{r} & \mathbf{0}
\end{array}\right]\left[\begin{array}{c}
\mathbf{U}^{\xi e}\\
\mathbf{U}^{\xi n}\\
\mathbf{f}^{\xi}
\end{array}\right]=\left[\begin{array}{c}
\mathbf{B}^{e}\\
\mathbf{B}^{n}\\
\mathbf{D}
\end{array}\right],
\end{equation}
with
\begin{align*}
A_{ia,jb}^{ee} & =\delta_{ab}\delta_{ij}(\epsilon_{i}-\epsilon_{a})-(2(ai|bj)+2K_{ai,bj}^{xc}),\\
A_{ia,J_{n}B_{n}}^{en} & =2(Z_{n}(ai|B_{n}J_{n})-K_{ai,B_{n}J_{n}}^{xc}),\\
A_{I_{n}A_{n},jb}^{ne} & =2(Z_{n}(A_{n}I_{n}|bj)-K_{A_{n}I_{n},bj}^{xc}),\\
A_{I_{n}A_{n},J_{n'}B_{n'}}^{nn} & =\delta_{A_{n}B_{n'}}\delta_{I_{n}J_{n'}}(\epsilon_{I_{n}}-\epsilon_{A_{n}})-2(Z_{n}Z_{n'}(A_{n}I_{n}|B_{n'}J_{n'})+K_{A_{n}I_{n},J_{n'}B_{n'}}^{xc}),\\
\mathbf{r}_{I_{n}A_{n},J_{n'}B_{n'}}^{nn} & =\delta_{A_{n}B_{n'}}\delta_{I_{n}J_{n'}}\mathbf{r}_{A_{n}I_{n}},\\
B_{ia}^{e} & =F_{ai}^{\xi}-S_{ai}^{\xi}\epsilon_{i}-\sum_{j}\sum_{k}((ai|kj)+K_{ai,jk}^{xc})S_{jk}^{\xi}+\sum_{n}(Z_{n}(ai|J_{n}J_{n})-K_{ai,J_{n}J_{n}}^{xc})S_{J_{n}J_{n}}^{\xi},\\
B_{I_{n}A_{n}}^{n} & =F_{A_{n}I_{n}}^{\xi}-S_{A_{n}I_{n}}^{\xi}\epsilon_{I_{n}}+\sum_{j}\sum_{k}(Z_{n}(A_{n}I_{n}|kj)-K_{A_{n}I_{n},jk}^{xc})S_{jk}^{\xi},\\
\mathbf{D}_{n} & =\frac{1}{2}(\frac{\partial\mathbf{R}_{n}}{\partial\xi}-\mathbf{r}_{I_{n}I_{n}}^{\xi}+\mathbf{R}_{n}S_{I_{n}I_{n}}^{\xi}).
\end{align*}
When $\xi$ is the expectation position of a particular quantum nucleus
$\xi=\mathbf{R}_{n_{0}}$, the matrix elements for $\mathbf{D}_{n}$
becomes
\begin{equation}
\mathbf{D}_{n}=\frac{1}{2}(\delta_{n,n_{0}}-\mathbf{r}_{I_{n}I_{n}}^{\mathbf{R}_{n_{0}}}+\mathbf{R}_{n}S_{I_{n}I_{n}}^{\mathbf{R}_{n_{0}}}).
\end{equation}
Solving cNEO-CPKS gives rise to $\mathbf{U}^{\xi}$ matrices, which
are used to evaluate the second-order derivatives in Eq. \ref{eq:energy_2de}.

\section{Computational details}

We implemented cNEO-CPKS equations and analytic Hessian of cNEO-DFT
in an in-house version of PySCF package.\citep{Sun2018,Sun2020} The
analytic Hessian results agree with those obtained with finite difference
(see Supporting Information for details), indicating the correct equation
derivation and code implementation. In all the subsequent calculations,
if not specially specified, the B3LYP functional\citep{Becke1988,Lee1988,Becke1993}
which is good in predicting molecular vibrational frequencies\citep{Scott1996}
is used for the electronic exchange-correlation. The cc-pVTZ basis
set\citep{Dunning1989} is adopted for electrons in regular diatomic
and polyatomic molecules, and the aug-cc-pVTZ basis set is used for
systems with a shared proton. In cNEO-DFT calculations, all quantum
nuclei, no matter bosons or fermions, are treated as distinguishable
particles at the Hartree level with the self-interaction excluded.
As a proof of principle, the current work does not include electron-nuclei
correlations or nuclei-nuclei correlations, and those will be left
for future studies. The even-tempered Gaussian basis\citep{Bardo1974}
is used for quantum nuclei. Specifically, the $8s8p8d$ basis set
with $\alpha=2\sqrt{2}$ and $\beta=\sqrt{2}$ is employed for protons,
and the $12s12p12d$ basis set is used for all the remaining nuclei
with $\beta=\sqrt{3}$ and $\alpha=4\sqrt{2}$, $12\sqrt{2}$, $14\sqrt{2}$,
$16\sqrt{2}$, and $18\sqrt{2}$ for D, C, N, O, and F, respectively.\citep{Xu2020a}
For each method, the stationary-point geometries are optimized using
the analytic gradient\citep{Xu2020a} by the Broyden-- Fletcher--Goldfarb--Shanno
(BFGS) algorithm implemented in the Atomic Simulation Environment
(ASE) package,\citep{Larsen2017} and the analytic Hessian calculations
and vibrational analysis are performed at the optimized geometries.
The vibrational frequencies for a series of small molecules are calculated
with both full-quantum cNEO-DFT and cNEO-DFT with only key protons
treated quantum-mechanically. For comparison, conventional DFT harmonic
vibrational analysis and DFT-VPT2 calculations\citep{Barone2005}
are performed with Gaussian 16.\citep{g16} Vibrational frequencies
by NEO-DFT(V) from Refs. \citenum{Yang2019} and \citenum{Culpitt2019}
are also compared. A very tight convergence criterion\citep{g16}
is used to optimize the geometry for DFT-VPT2 calculations as recommended.\citep{Barone2005}

\section{Results and discussions}

\subsection{Vibrational frequencies of diatomic molecules}

The conventional NEO calculation requires at least two atoms to be
treated classically to avoid the challenges from translational and
rotational symmetry,\citep{Iordanov2003} and therefore faces challenges
in describing monoatomic and diatomic molecules. In cNEO theory, the
expectation position for the quantum nuclei fixes the molecular frame,
making it possible to handle monoatomic and diatomic molecules. We
calculated the molecular vibrational frequencies of 7 simple diatomic
molecules by diagonalizing the cNEO-DFT Hessian matrix. The results
are listed in Table 1 together with experimental values and computational
results from diagonalizing DFT Hessian matrix (DFT-Harmonic) and from
DFT-VPT2. Compared to the experimental values, the full-quantum cNEO-DFT
underestimates the vibrational frequencies of \ce{H2}/HD/\ce{D2}
and HF, and overestimates those of \ce{F2} and \ce{N2}. The mean
signed error (MSE) is -3.1 cm$^{-1}$ and the mean unsigned error
(MUE) is 101.1 cm$^{-1}$. In contrast, DFT harmonic calculations
always overestimate the vibrational frequencies of these molecules,
and the MSE and MUE are both 165.9 cm$^{-1}$. DFT-VPT2 performs best
with an MSE of 43.4 cm$^{-1}$ and an MUE of 57.5 cm$^{-1}$.

Comparing \ce{H2} to HD, the deuterization drops the experimental
vibrations frequencies by $\sim$530 cm$^{-1}$. Further deuterization
to \ce{D2} reduces the frequency by another $\sim$630 cm$^{-1}$.
This isotope effect is well captured by cNEO-DFT. With cNEO-DFT, the
frequency difference between \ce{H2} and HD is 500 cm$^{-1}$ and
the difference between HD and \ce{D2} is 610 cm$^{-1}$. DFT-harmonic
is less accurate and predicts 590 cm$^{-1}$ and 700 cm$^{-1}$ drops
for the deuterations. DFT-VPT2 is most accurate and predicts the frequency
drops to be 530 cm$^{-1}$ and 640 cm$^{-1}$, respectively.

\begin{table}
\caption{Vibrational frequencies of 7 diatomic molecules (in cm$^{-1}$)}

\begin{tabular}{cccccc}
\hline 
Molecule & Vibrational mode & Experiment$^{a}$ & Full-quantum cNEO-DFT & DFT-harmonic & DFT-VPT2\tabularnewline
\hline 
\ce{H2} & H-H stretch & 4161.2 & 4045.2 & 4419.5 & 4187.2\tabularnewline
HD & H-D stretch & 3632.2 & 3545.5 & 3827.9 & 3653.2\tabularnewline
\ce{D2} & D-D stretch & 2993.7 & 2930.1 & 3126.3 & 3010.4\tabularnewline
HF & F-H stretch & 3961.4 & 3914.9 & 4092.3 & 3919.2\tabularnewline
DF & F-D stretch & - & 2869.6 & 2966.7 & 2875.9\tabularnewline
\ce{F2} & F-F stretch & 893.9 & 1055.7 & 1052.0 & 1038.3\tabularnewline
\ce{N2} & N-N stretch & 2329.9 & 2462.2 & 2450.0 & 2424.8\tabularnewline
\hline 
MSE &  &  & -3.1 & 165.9 & 43.4\tabularnewline
MUE &  &  & 101.1 & 165.9 & 57.5\tabularnewline
\hline 
\end{tabular}
\raggedright{}a. From the National Institute of Standards and Technology
(NIST) websites
\end{table}

\subsection{Vibrational frequencies of polyatomic molecules}

The vibrational frequencies of 11 simple polyatomic molecules are
presented in Table 2. We performed two kinds of cNEO-DFT calculations.
One is the full-quantum version, in which all atoms are treated quantum
mechanically. The other one only treats hydrogen atoms quantum mechanically.
The vibrational frequencies calculated by harmonic DFT, DFT-VPT2,
and NEO-DFT(V) are also shown for comparison.

All methods perform significantly better in polyatomic molecules than
in diatomic molecules. The MSE and MUE for full-quantum cNEO-DFT is
4.9 cm$^{-1}$ and 28.8 cm$^{-1}$, respectively. It is similar to
cNEO-DFT with only hydrogen atoms treated quantum mechanically. The
overall performance of cNEO-DFT is comparable to that of DFT-VPT2
in these polyatomic molecules. The MUE of DFT-VPT2 is 26.2 cm$^{-1}$,
which is on average 2.6 cm$^{-1}$ more accurate than full-quantum
cNEO-DFT. Harmonic DFT is still the least accurate method with 66.1
cm$^{-1}$ MSE and 66.9 cm$^{-1}$ MUE, which almost double those
of cNEO-DFT and DFT-VPT2. Previously, NEO-DFT(V), a method that combines
NEO-DFT Hessian with NEO-TDDFT, was developed to incorporate nuclear
quantum effects in the molecular vibrational analysis.\citep{Yang2019}
It greatly outperforms DFT harmonic calculations, especially for vibrations
with significant anharmonicity. However, with a 21.3 cm$^{-1}$ MSE
and a 39.5 cm$^{-1}$ MUE, it is not accurate as cNEO-DFT. Furthermore,
cNEO-DFT is more efficient than NEO-DFT(V) since NEO-DFT(V) requires
an additional multicomponent TDDFT calculation with large basis sets.\citep{Culpitt2019a}

The calculated vibrational frequencies are plotted against the experimental
values in Fig. 1. Harmonic DFT calculations tend to overestimate vibrational
frequencies, especially for the X-H stretch vibration modes whose
frequencies are above 3000 cm$^{-1}$. These modes are traditionally
known to have large anharmonicity. NEO-DFT(V) faces challenges in
describing some low-frequency X-H bend modes, in particular the bend
modes in HNC, \ce{C2H2} and \ce{H2O2}. In contrast, cNEO-DFT can
give a better description for both X-H stretch and X-H bend modes.
In fact, if we only consider the X-H modes, which play a more important
role in many chemical reactions, the MUEs of full-quantum cNEO-DFT
and DFT-VPT2 are the same. This result is surprising because the frequencies
by cNEO-DFT come from a harmonic treatment, while the X-H modes were
considered to have significant anharmonicity. The reason for this
good performance of harmonic cNEO-DFT may be the inclusion of nuclear
quantum effects, especially the delocalized nuclear orbital picture.
Previously, a similar redshift on vibrational frequencies has been
found when comparing the vibrational spectra from classical molecular
dynamics and from path-integral molecular dynamics (PIMD)\citep{Habershon2008,Kaczmarek2009,Calvo2014}
Both classical molecular dynamics and PIMD include anharmonic effects
in the simulation. However, PIMD, which includes the nuclear quantum
effects in the path-integral formulation, gives lower and more accurate
vibrational frequencies. All these facts suggest that the nuclear
quantum effects might play a more important role than anharmonic effects
in the accurate description of molecular vibrations.

\begin{table}
\caption{Vibrational frequencies of 11 simple polyatomic molecules (in cm$^{-1}$)}

\tiny

\begin{tabular*}{1\textwidth}{@{\extracolsep{\fill}}cccccccc}
\hline 
Molecule & Vibrational mode & Experiment$^{a}$ & Full-quantum cNEO-DFT & cNEO-DFT quantum H & DFT-harmonic & DFT-VPT2 & NEO-DFT(V)$^{b}$\tabularnewline
\hline 
\multirow{3}{*}{HCN} & C-H stretch & 3312.0 & 3317.9 & 3308.4 & 3450.0 & 3322.1 & 3317\tabularnewline
 & C-N stretch & 2089.0 & 2200.6 & 2190.0 & 2201.4 & 2175.3 & 2191\tabularnewline
 & C-H bend & 712.0 & 740.3 & 736.7 & 762.1 & 750.9 & 789\tabularnewline
\hline 
\multirow{3}{*}{HNC} & N-H stretch & 3652.9 & 3657.0 & 3630.4 & 3806.9 & 3642.1 & 3645\tabularnewline
 & N-C stretch & 2029.0 & 2107.5 & 2100.0 & 2106.2 & 2072.3 & 2100\tabularnewline
 & N-H bend & 477.0 & 466.1 & 457.3 & 472.5 & 468.4 & 568\tabularnewline
\hline 
\multirow{6}{*}{HCFO} & C-H stretch & 2981 & 2944.4 & 2940.6 & 3080.3 & 2941.9 & 2947\tabularnewline
 & C-O stretch & 1837 & 1898.6 & 1888.4 & 1893.4 & 1861.4 & 1885\tabularnewline
 & C-H bend & 1342 & 1318.6 & 1311.9 & 1370.3 & 1340.2 & 1329\tabularnewline
 & C-F stretch & 1065 & 1073.7 & 1065.0 & 1072.9 & 1048.6 & 1075\tabularnewline
 & C-H out-of plane bend & 1011 & 1016.1 & 1008.3 & 1035.8 & 1019.4 & 1061\tabularnewline
 & FCO bend & 663 & 669.9 & 665.2 & 666.5 & 659.2 & 665\tabularnewline
\hline 
\multirow{6}{*}{\ce{HCF3}} & C-H stretch & 3035 & 2973.2 & 2968.7 & 3117.1 & 2999.3 & 2988\tabularnewline
 & C-H bend & 1376 & 1346.2 & 1342.4 & 1393.8 & 1361.6 & 1353\tabularnewline
 & CF asymmetric stretch & 1152 & 1143.8 & 1135.1 & 1145.3 & 1118.4 & 1134\tabularnewline
 & CF symmetric stretch & 1137 & 1136.9 & 1130.2 & 1136.2 & 1118.0 & 1128\tabularnewline
 & CF simultaneous bend & 700 & 699.9 & 695.7 & 695.9 & 688.2 & 693\tabularnewline
 & FCF scissor & 508 & 506.1 & 502.8 & 503.2 & 497.4 & 501\tabularnewline
\hline 
\multirow{5}{*}{\ce{C2H2}} & C-H stretch & 3374.0 & 3391.3 & 3383.9 & 3518.4 & 3389.5 & 3378\tabularnewline
 & C-H stretch & 3289.0 & 3268.8 & 3271.1 & 3414.9 & 3293.5 & 3263\tabularnewline
 & C-C stretch & 1974.0 & 2055.8 & 2054.7 & 2071.7 & 2040.5 & 2047\tabularnewline
 & C-H bend & 730.0 & 737.0 & 739.4 & 767.9 & 754.2 & 786\tabularnewline
 & C-H bend & 612.0 & 647.9 & 637.2 & 653.2 & 670.8 & 727\tabularnewline
\hline 
\multirow{6}{*}{\ce{H2CO}} & \ce{CH2} a-stretch & 2843.0 & 2786.3 & 2780.7 & 2933.6 & 2683.2 & 2772\tabularnewline
 & \ce{CH2} s-stretch & 2782.0 & 2743.9 & 2733.0 & 2879.1 & 2728.4 & 2724\tabularnewline
 & C-O stretch & 1746.0 & 1823.5 & 1810.6 & 1823.1 & 1797.1 & 1812\tabularnewline
 & \ce{CH2} scissors & 1500.0 & 1475.9 & 1475.7 & 1536.0 & 1503.2 & 1477\tabularnewline
 & \ce{CH2} rock & 1249.0 & 1228.7 & 1225.1 & 1268.1 & 1247.5 & 1254\tabularnewline
 & \ce{CH2} wag & 1167.0 & 1168.6 & 1163.8 & 1202.2 & 1183.7 & 1190\tabularnewline
\hline 
\multirow{6}{*}{\ce{H2O2}} & O-H a-stretch & 3608.0 & 3620.2 & 3597.9 & 3761.2 & 3581.0 & 3599\tabularnewline
 & O-H s-stretch & 3599.0 & 3617.5 & 3595.3 & 3759.8 & 3583.2 & 3596\tabularnewline
 & O-H s-bend & 1402.0 & 1381.1 & 1375.5 & 1439.3 & 1400.9 & 1425\tabularnewline
 & O-H a-bend & 1266.0 & 1260.5 & 1254.1 & 1323.3 & 1274.0 & 1314\tabularnewline
 & O-O stretch & 877.0 & 953.0 & 945.5 & 953.3 & 928.2 & 957\tabularnewline
 & HOOH torsion & 371.0 & 333.3 & 328.7 & 374.9 & 324.9 & 523\tabularnewline
\hline 
\multirow{6}{*}{\ce{H2NF}} & asymmetric NH stretch & 3346 & 3350.0 & 3326.7 & 3499.9 & 3316.0 & 3336\tabularnewline
 & symmetric NH stretch & 3234 & 3258.8 & 3232.5 & 3406.2 & 3247.0 & 3241\tabularnewline
 & \ce{NH2} scissor & 1564 & 1540. 8 & 1536.0 & 1620.5 & 1565.2 & 1556\tabularnewline
 & \ce{NH2} wag/rock & 1241 & 1287.9 & 1284.9 & 1337.0 & 1302.4 & 1310\tabularnewline
 & \ce{NH2} wag/rock & 1233 & 1192.9 & 1208.1 & 1259.7 & 1217.7 & 1257\tabularnewline
 & NF stretch & 891 & 931.5 & 923.1 & 938.5 & 913.6 & 936\tabularnewline
\hline 
\multirow{3}{*}{\ce{H2O}} & O-H a-stretch & 3756.0 & 3756.9 & 3730.6 & 3901.3 & 3715.5 & -\tabularnewline
 & O-H s-stretch & 3657.0 & 3658.6 & 3633.3 & 3800.9 & 3628.2 & -\tabularnewline
 & O-H bend & 1595.0 & 1533.9 & 1543.4 & 1639.5 & 1586.9 & -\tabularnewline
\hline 
\multirow{9}{*}{HCOOH} & O-H stretch & 3570.0 & 3549.1 & 3524.8 & 3722.1 & 3534.6 & -\tabularnewline
 & C-H stretch & 2943.0 & 2901.8 & 2896.5 & 3041.9 & 2891.5 & -\tabularnewline
 & C=O stretch & 1770.0 & \noun{1826.0} & 1813.4 & \noun{1826.6} & 1793.7 & -\tabularnewline
 & C-H bend & 1387.0 & 1351.5 & 1347.6 & 1406.1 & 1378.5 & -\tabularnewline
 & O-H bend & 1229.0 & 1272.8 & 1269.0 & 1306.0 & 1225.7 & -\tabularnewline
 & C-O stretch & 1105.0 & 1116.2 & 1108.8 & 1125.1 & 1093.0 & -\tabularnewline
 & C-H bend & 1033.0 & 1035.6 & 1028.1 & 1056.1 & 1036.2 & -\tabularnewline
 & torsion & 638.0 & 672.6 & 670.2 & 683.4 & 647.3 & -\tabularnewline
 & OCO bend & 625.0 & 629.1 & 625.6 & 629.8 & 623.2 & -\tabularnewline
\hline 
MSE &  &  & 4.9 & -2.6 & 66.1 & -1.9 & 21.3\tabularnewline
MUE &  &  & 28.8 & 29.4 & 66.9 & 26.2 & 39.5\tabularnewline
MUE(X-H modes) &  &  & 25.2 & 27.6 & 81.1 & 25.2 & 36.1\tabularnewline
\hline 
\end{tabular*}
\begin{raggedright}
a. Experimental data are from the National Institute of Standards
and Technology (NIST) websites.
\par\end{raggedright}
\raggedright{}b. NEO-DFT(V) results with B3LYP and epc17-2 functionals
are directly taken from Refs. \citenum{Yang2019} and \citenum{Culpitt2019}.
\end{table}

\begin{figure}
\includegraphics{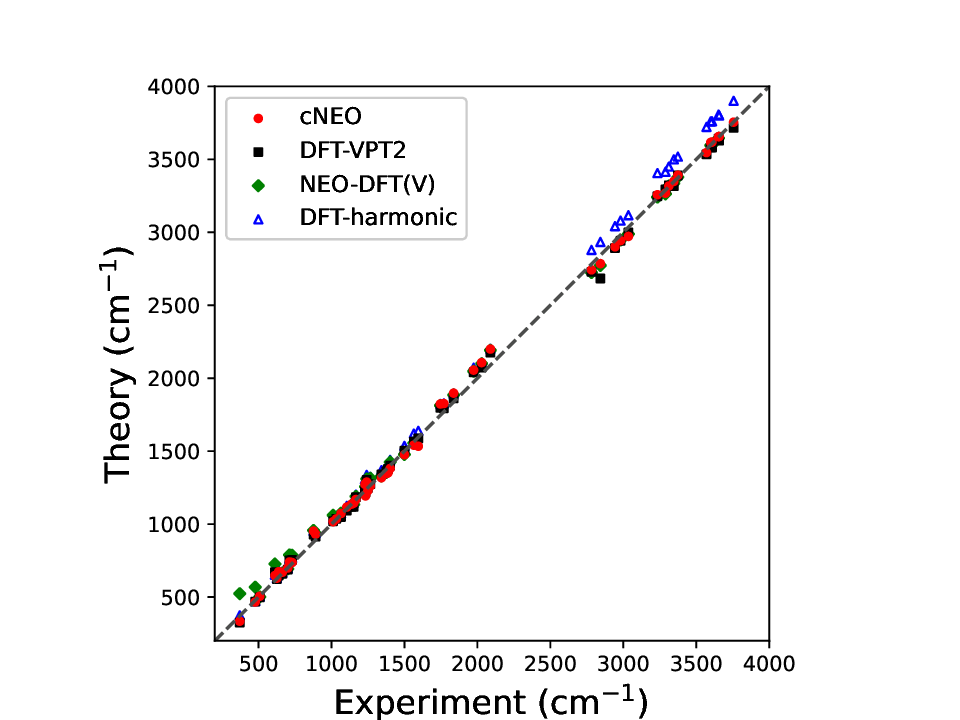}

\caption{Experimental and theoretical vibrational frequencies for molecules
in Table 2.}
\end{figure}

\subsection{Vibrational frequencies of proton transfer modes in systems with
a shared proton}

Proton transfer through hydrogen bonds is crucial for understanding
dynamic properties in many chemical and biological systems. The nuclear
quantum effect is believed to play an important role during the proton
transfer process.\citep{Tuckerman1997,Marx1999} Here we apply cNEO-DFT
Hessian to 4 simple systems with shared protons, including \ce{FHF-},
\ce{H3O2-}, \ce{H5O2+} and \ce{N2H7+}.

From geometry optimization, both DFT and cNEO-DFT predict the shared
proton for \ce{FHF-} and \ce{H5O2+} to be in the middle of the two
heavy atoms. As for \ce{H3O2-} and \ce{N2H7+}, DFT gives a double-well
potential energy surface and predicts the proton to be either on the
left or on the right, whereas cNEO-DFT still predicts the proton to
be in the middle. This is consistent with the results from previous
literature and the reason is that when the proton is treated quantum
mechanically, the vibrational zero-point energy washes out the low
proton transfer barrier.\citep{Asmis2007,Yang2008,Kaledin2009}

The experimental and computational vibrational frequencies for the
proton transfer modes are presented in Table 3. Full-quantum cNEO-DFT
overestimates the frequencies of \ce{FHF-}, \ce{H3O2-} and \ce{H5O2+}
by 50-300 cm$^{-1}$ and accurately predict that of \ce{N2H7+}.
In contrast, DFT harmonic calculations severely overestimate the frequency
of \ce{N2H7+} but seem to give acceptable results for the first three
molecules. However, the bad performance of DFT-VPT2 and the large
discrepancy between DFT and DFT-VPT2 show the failure of the perturbative
treatment for these low-frequency modes and indicate that the DFT
potential energy surfaces are not really reliable in these systems.
In addition, the vibrational frequencies for these modes by DFT-VPT2
can be very sensitive to the choice of electronic exchange-correlation
functionals. We compare the results by B3LYP and PBE0 in Table 3.
A huge difference can be found for the anharmonic frequencies of the
proton transfer mode of \ce{H3O2-} and \ce{N2H7+} by DFT-VPT2. In
contrast, cNEO-DFT is less sensitive to functional choice with B3LYP
and PBE0 functionals giving similar results within 150 cm$^{-1}$.
Previously anharmonic frequencies by VPT2 have also been found to
be very sensitive to the choice of underlying computational methods
and basis sets.\citep{HansonHeine2012,Jacobsen2013} For example,
for \ce{FHF-}, VPT2 based on the coupled-cluster theory can give
a much more accurate result (1313 cm$^{-1}$) than DFT-VPT2,\citep{Hirata2008}
and for \ce{N2H7+}, MP2-VPT2 predicts a frequency of 485 cm$^{-1}$,\citep{Wang2016a}
which is also much closer to the experimental value than DFT-VPT2.
Therefore, although VPT2 can give improved results with a more accurate
underlying theory, cNEO-DFT is able to give reasonably good results
based on commonly used density functionals with a much lower computational
cost. This makes cNEO-DFT a promising method for the quantum description
of protons in large chemical systems in future studies.

\begin{table}
\caption{Vibrational frequencies of \ce{FHF-}, \ce{H3O2-}, \ce{H5O2+} and
\ce{N2H7+} (in cm$^{-1}$)}

\begin{tabular}{ccccccccccc}
\hline 
\multirow{2}{*}{Molecule} & \multirow{2}{*}{Ptoton transfer mode} & \multirow{2}{*}{Experiment} & \multicolumn{2}{c}{Full-quantum cNEO-DFT} &  & \multicolumn{2}{c}{DFT-harmonic} &  & \multicolumn{2}{c}{DFT-VPT2}\tabularnewline
\cline{4-5} \cline{5-5} \cline{7-8} \cline{8-8} \cline{10-11} \cline{11-11} 
 &  &  & B3LYP & PBE0 &  & B3LYP & PBE0 &  & B3LYP & PBE0\tabularnewline
\hline 
\multirow{1}{*}{\ce{FHF-}} & F-H stretch & 1331.2$^{a}$ & 1584.5 & 1679.1 &  & 1286.3 & 1365.7 &  & 1550.1 & 1574.8\tabularnewline
\ce{H3O2-} & O-H stretch & 697$^{b}$ & 756.8 & 941.7 &  & 316.3 & 299.7 &  & 144.7i & 1270.6i\tabularnewline
\ce{H5O2+} & O-H stretch & 1085$^{c}$ & 1181.0 & 1254.7 &  & 925.3 & 982.7 &  & 1406.6 & 1390.4\tabularnewline
\ce{N2H7+} & N-H stretch & 374$^{d}$ & 324.0 & 405.3 &  & 1727.9 & 1731.1 &  & 182.2 & 37.4\tabularnewline
\hline 
\end{tabular}
\begin{raggedright}
a. From Ref. \citenum{Kawaguchi1987}
\par\end{raggedright}
\begin{raggedright}
b. From Ref. \citenum{Diken2005}
\par\end{raggedright}
\begin{raggedright}
c. From Ref. \citenum{Headrick2005}
\par\end{raggedright}
\raggedright{}d. From Ref. \citenum{Yang2008}
\end{table}

\section{Conclusion}

In this work, we derived and implemented the first-order CPKS equation
and the analytic Hessian for cNEO-DFT. The cNEO-CPKS equation incorporates
the constraints on the nuclear expectation positions and solves the
responses of electronic and nuclear orbitals to the changes of quantum
nuclear expectation positions as well as classical nuclear positions.
These responses are used in the calculation of cNEO-DFT analytic Hessian.
Except diatomic molecules, cNEO-DFT are comparable to DFT-VPT2 in
the description of molecular vibrational frequencies in polyatomic
molecules. The incorporation of nuclear quantum effects in the energy
surface enables cNEO-DFT to well describe many vibrational modes that
were previously considered anharmonic with a simple harmonic treatment.
These modes include the stretch and bend modes of X-H bonds, which
are important in describing many chemical reactions. Compared to NEO-DFT(V),
cNEO-DFT is also much more accurate and more efficient. For systems
with shared protons, cNEO-DFT can be a more reliable method than DFT-VPT2
because the energy surface by cNEO-DFT includes nuclear quantum effects
and is more reliable than the DFT potential energy surface for these
challenging systems. These results indicate the good quality of the
cNEO-DFT energy surface, and since Hessian can be used to characterize
local minima and saddle points, this work makes cNEO-DFT a promising
and viable method to include nuclear quantum effects in the study
of chemical reactions through transition-state theory or molecular
dynamics simulations in the future.

\section*{Supplementary Material}

See the supplementary material for detailed derivation for the analytic
Hessian of cNEO-DFT and a comparison of analytic and finite-difference
results for molecular vibrational frequencies.
\begin{acknowledgments}
We are grateful for the support and funding from the University of
Wisconsin via the Wisconsin Alumni Research Foundation. We also thank
Dr. Zehua Chen for helpful discussions.
\end{acknowledgments}

\section*{data avalibility}

The data that support the findings of this study are available within
the article and its supplementary material.

\bibliographystyle{apsrev4-1}
%\bibliography{dft}

\end{document}